\documentclass[letterpaper,english,aps,prl,twocolumn,groupedaddress]{revtex4}
\usepackage[T1]{fontenc}
\usepackage[latin1]{inputenc}
\usepackage{graphicx}
\usepackage{amssymb}

\makeatletter

\usepackage{babel}
\makeatother
\begin{document}

\title{Topological Analysis of Linear Polymer Melts}

\author{Christos Tzoumanekas}

\email{tzoum@central.ntua.gr}

\author{Doros N. Theodorou}

\email{doros@chemeng.ntua.gr}

\affiliation{Department of Materials Science and Engineering, School of Chemical
Engineering, National Technical University of Athens, Zografou Campus,
15780 Athens, Greece}

\affiliation{Dutch Polymer Institute (DPI), The Netherlands}

\begin{abstract}
We introduce an algorithm for the reduction of computer generated
atomistic polymer samples to networks of primitive paths. By examining
network ensembles of Polyethylene and cis-1,4 Polybutadiene melts,
we quantify the underlying topologies through the radial distribution
function of entanglements and the distribution of the number of monomers
between entanglements. A suitable scaling of acquired data leads to
a unifying microscopic topological description of both melts.
\end{abstract}



\date{\today{}}

\maketitle
Polymer chains cannot cross each other. A successful conceptual framework
embodying this principle at the molecular level is offered by the
tube model \cite{DeGennes,Doi}. The tube model postulates that the
mutual uncrossability of polymer chains generates topological constraints
(TCs), referred to as entanglements, which effectively restrict individual
chain conformations in a curvilinear tube-like region enclosing each
chain. Lateral chain motion is confined to the length scale of the
tube diameter. Large-scale motion is promoted via reptation \cite{DeGennes},
an effective one-dimensional diffusion of a chain along its tube axis.
The latter provides a coarse-grained representation that characterizes
the chain topology and is called the Primitive Path (PP). Edwards
\cite{Edwards} regarded the PP as the shortest path constructed by
keeping the chain ends fixed while continuously tightening (shrinking)
the chain contour, so that the resulting path has the same topology
relative to other chains as the chain itself \cite{Doi,Edwards}. Applying
this construction for all chains \cite{Rubi,Ever} leads to a coarse-grained
picture of a polymer melt that uncovers its large-scale \emph{topological
substructure}. The latter is conceived as a network of entangled PPs
underlying the melt structure \cite{Masu}. The tube diameter $d$ is assumed uniform
and corresponds to the mesh length of this network \cite{Doi}.

Entanglements dominate the rheological and dynamical properties of
large-molecular weight polymer melts \cite{Doi}, as well as the ultimate
mechanical properties in the glassy state of these systems \cite{Kramer}.
Their topological nature makes their direct experimental study difficult
and their microscopic definition elusive \cite{Richter}. In this letter,
we present \emph{microscopically determined distributions} that describe in a 
statistical manner the topological state of flexible polymer melts. 
We utilize an algorithm, referred to as CReTA (Contour Reduction Topological
Analysis), which is capable of reducing the atomistic configuration
of a computational polymer sample to a network of corresponding
PPs. Nodal points of the reduced network correspond to TCs 
experienced by individual chains, while network connectivity defines the underlying 
melt topology. 
Topological measures extracted from networks of entangled Polyethylene (PE) and cis-1,4 Polybutadiene
(PB) melts are compared against corresponding experimental data. Quantitative agreement establishes
these networks as meaningful, well defined, structural representations
of the underlying melt topologies. 

Mesoscopic views where a polymer melt is represented as an entanglement network (rubber analogy) 
have long been used as conceptual abstractions for the development of 
phenomenological models (transient network theory), and analytical theories (tube model) in polymer physics.
These views are guided from the rubber-like response of melts in viscoelastic experiments (rubbery plateau in the
relaxation modulus), and the Random Walk (RW) statistics underlying the structure of flexible polymers. 
RW-like chain configurations, above some length scale show a self-similarity, to which 
universalities and scaling laws detactable in physical properties can be traced. 
Here, by examining networks generated from well packed RW-like chain configurations of chemically 
different polymer melts,
we characterize the melt topology through \emph{distributions} which, when scaled accordingly, 
unveil a \emph{unifying topological description}. The reported statistical properties of 
entanglement networks constitute a \emph{missing link} for the construction of mesoscopic 
simulation models of polymer melts  \cite{Masu} and glasses \cite{Terzis} suitable for the prediction of rheological 
and large-deformation mechanical properties. One can use PP statistics  
in reverse-engineering fashion to construct entanglement networks obeying the mentioned 
distributions, for any flexible polymer. Microscopically determined PPs offer
a promising basis \cite{Ever,Larson,Kroger} for investigating the general picture of entanglements invoked by
the tube model.

Following Edwards's perspective and as proposed in
\cite{Rubi,Ever}, CReTA provides a solution to
the following geometric problem \cite{Kroger}: 
`Given a set of uncrossable curved lines in space, 
reduce continuously their contour lengths keeping their ends fixed, 
until they become sets of rectilinear segments (entanglement strands) 
coming together at nodal points (entanglements)'. For each polymer we examine
an ensemble of statistically independent atomistic samples. The ensembles have
been thermodynamically and topologically equilibrated at all length
scales by previous Monte Carlo simulations \cite{Kar,Gest} employing
chain-connectivity altering moves which alter the system topology. 
Through CReTA, atomistic configurations are transformed to
networks, which are then analyzed to extract topological properties.

CReTA implements random aligning string moves and hardcore interactions.
CH$_{x}$ monomers are treated as united-atom hard spheres of diameter
$\sigma$. Chains are represented as series of fused spheres, since
in atomistic polymer models the skeletal bond distance ($\sim1.5\textrm{Å}$)
is lower than $\sigma$ ($\sim3.5\textrm{Å}$). Periodic boundary
conditions apply in all cubic sample directions. A string is defined
as a set of $m\geq1$ consecutive chain atoms. On each move, a string
is randomly chosen and the string atoms are displaced to corresponding
equidistant points on the straight line segment joining the atoms
on either side of the string. Chain ends are fixed throughout the
process. To avoid chain crossing and preserve system topology, moves
leading to overlaps between any string atom with any other atom belonging
to a different chain are rejected. This is sufficient for preventing
crossing when chains consist of fused spheres. $m$ is randomly chosen
from an interval of $\Delta m\sim10$ width. Accepted moves result
in simultaneous reduction of chain contours and progressive shrinkage
of unentangled loops to straight strands composed of fused spheres.
Rejected moves stem from either mutually blocked chain parts, which
will lead to entanglements, or from contacts between unentangled loops,
which will eventually disappear. To accelerate chain tightening we
increase string sizes by gradually increasing $\Delta m$. When chain
contour lengths are no longer diminishing, chain thickness is reduced
by decreasing $\sigma$ and the process starts anew. This aids in
tightening meshed unentangled loops, which, although temporarily blocked,
do not represent true TCs. Upon decreasing $\sigma$, to preserve
the fused sphere sequence of chains we place an auxiliary atom between
successive skeletal atoms that have lost contact. Auxiliary atoms
are excluded from latter analysis. The whole procedure terminates
when a predefined $\sigma_{f}\sim0.5\textrm{Å}$ is reached. At this
point an underlying network structure of interpenetrating `zigzag'
shaped PPs has been revealed \cite{foot1}. Further $\sigma$ reduction
to attain the infinitely-thin-continuous line limit would be time
consuming and superfluous. CReTA differs from recent similar approaches.
In \cite{Ever}, contour reduction is achieved through the minimization
of an elastic energy which does not lead to well defined networks
of `kinky' PPs where nodal points can be specified. Kr\"oger's approach \cite{Kroger} is 
also geometric and capable of generating networks 
but, unlike CReTA, cannot track a monomer in the system 
both in the melt and network configurations. The latter
feature offers great potential for studying hidden topological
correlations in space, time, and along the chains, in close connection 
with the melt structure and dynamics.
The $P(n)$ distribution presented later is an example of this.

\begin{figure}
\begin{center}\includegraphics[%
  clip,
  scale=0.38]{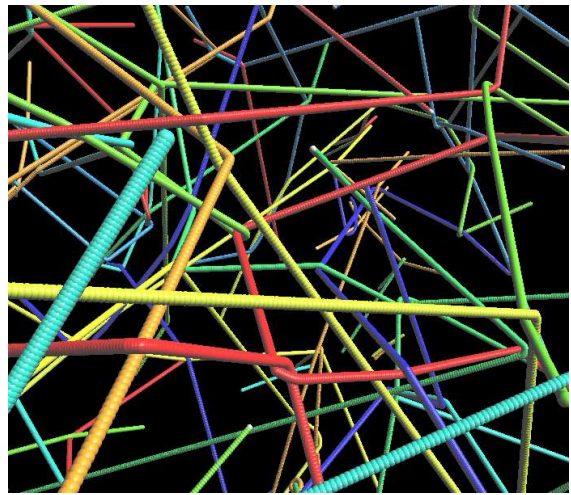}\end{center}

\caption{\label{cap:sample}(color online). PP network view of a PE melt.}
\end{figure}

As shown in Fig. \ref{cap:sample}, the resulting PPs consist of short
pairwise blocked chain sections connected by straight strands of fused
atoms. By examining blocked regions we resolve PP `contacts' to pairs of
neighboring monomers. These specify pairwise associated TCs in the
monomer sequence of their parent chains. In addition, they represent
effective spatial localization points of the TCs each chain is subjected
to. By convention, we refer to these points as entanglements.
Each PP is characterized by its ends and a set of entanglements with
specific spatial locations, monomer sequence indices, and pairwise
associations to other entanglements. The whole structure is reduced
to a network, with nodal points the entanglements, and edges the joining entanglement strands
(ESs). \textit{Topological analysis} reduces then to an examination
of network properties.  Entanglement spacing (network mesh) on the
PP contour is deduced both in monomer and length units by calculating
$\overline{N}_{{\rm ES}}$ and $\overline{d}_{{\rm ES}}$, the ensemble
average number of monomers and end-to-end length, respectively, of an
ES. Experimentally, entanglement spacing is mainly inferred either
from the ES monomer length $N_{e}$, measured from the plateau regime
of the dynamic shear modulus \cite{Fet}, or from the tube diameter
$d$ (entanglement distance) measured as an intermediate dynamic length
in neutron spin echo studies of the dynamic structure factor \cite{Richter}.
These are interpreted on the basis of the tube model, where PP conformations
are considered as RWs and an ES is identified with the PP
Kuhn segment \cite{Doi}. Here, the relevant quantities
are the PP Kuhn length $d=R^{2}/L$, and the number of monomers in
a PP Kuhn segment $N_{e}=NR^{2}/L^{2}$. $N$ denotes the chain average
number of monomers. $R^{2}$ and $L$ are the PP squared end-to-end
distance and contour length, respectively, calculated here as
ensemble averages. By this `Kuhn mapping' approach,
entanglement spacing emerges as the mesh spacing of an assumed uniform
PP network of interpenetrating random walks.

\begin{table}[!] 

\caption{\label{table}~Entanglement spacing in monomer ($\overline{N}_{\rm ES}$, $N_e$), and length 
($\overline{d}_{\rm ES}$, $d$) units of PP networks, and packing length $p$ of the  initial atomistic 
samples (equilibrated at $T=450$K (PE), $T=413$K (PB)).  $\overline{N}_{\rm ES}$, $\overline{d}_{\rm ES}$ 
refer to the natural network mesh. $N_e$, $d$ refer to the PP Kuhn mapping approach. Numbers in parentheses
are experimental values from \cite{Fet}, unless otherwise indicated. The calculated polymer densities are 
0.776(0.766)$^a$,  0.778(0.766)$^a$, 0.867(0.826)~$g/cm^{3}$; the polydispersity indices are 1.083, 1.000, 1.053; 
and the number of chains for each system studied are 16, 8, 24 (top to bottom).}

\begin{ruledtabular} 

\begin{tabular}{lcccccc}  

& $N$ & $\overline{N}_{\rm ES}$ & $\overline{d}_{\rm ES}$ (\AA)  & $N_e$ & $d$ (\AA) & $p$ (\AA) \\ 

\hline 

PE & 500 & 28.3 & 14.0 & 75.1(61.4) & 38.4(39.8)\footnotemark[2] & 1.53(1.69) \\ 

PE& 1000 & 29.1 & 14.1 & 74.2(61.4) & 36.6(39.8)\footnotemark[2] & 1.65(1.69) \\

PB& 1000 & 80.9 & 18.7 & 178.7(173.8) & 42.3(43.0)~ & 2.59(2.44)  \\

\end{tabular} 

\end{ruledtabular} 

\footnotetext[1]{Ref.~\onlinecite{Pearson}.} 

\footnotetext[2]{Ref.~\onlinecite{Richter}.}

\end{table} 

Table \ref{table} summarizes our results for  $\overline{N}_{\rm ES}$, $N_e$, 
$\overline{d}_{\rm ES}$, $d$, the packing length $p$ and
the density of the atomistic samples, along with
corresponding experimental values. $p$, an effective chain thickness that
controls coil packing, is defined as $1/(\rho_{ch}R^{2})$, where
$\rho_{ch}$ is the number density of chains. Volumetric, structural and conformational
predictions for both polymers \cite{Kar,Gest} are in excellent agreement with available 
experimental data. From
the additional $p$ data presented here, we conclude that the interplay
between large-scale chain conformation and monomer packing is nicely
captured by the atomistic ensembles. Turning to networks, PE samples
of $N=500,1000$ display practically the same mesh values, showing
that well entangled samples possess a \textit{quantitatively} similar
underlying topology, independently of chain length. The small differences
are attributed to different packing lengths. $\overline{N}_{{\rm ES}}$,
$\overline{d}_{{\rm ES}}$, the network natural mesh spacing quantities,
are much smaller than the corresponding PP Kuhn segment quantities
$N_{e}$, $d$, with $\overline{N}_{{\rm ES}}\sim0.4N_{e}$ for both
materials. This happens because of directional correlations between
ESs in the same PP that decay exponentially with ES separation.
That is, PP conformations are not RWs, and PP Kuhn segments
are not free of TCs. At the Kuhn scale, where PPs become RWs,
$N_{e},\: d$ for both polymers are in good agreement with experimental
data. If we consider our systems as rubber networks, however, the
small value of $\overline{N}_{{\rm ES}}$ implies a larger plateau
modulus $G_{o}$ than found experimentally. Interestingly, in recent
\cite{Masu} Brownian dynamics simulations of 3D PP networks, 
the linear viscoelastic response of PB solutions is reproduced
quantitatively only when an $N_{e}$ of about half the experimental
value is utilized, as found here for $\overline{N}_{{\rm ES}}$. The discrepancy 
is attributed \cite{Masu} to the affine as opposed
to phantom network model experimental estimate of $G_{o}$. In this
respect, our results are in favor of the phantom model, which for
a tetrafunctional network would estimate approximately half an $N_{e}$
for the same $G_{o}$ \cite{Masu}. 

\begin{figure}[t]
\begin{center}\includegraphics[%
  scale=0.42]{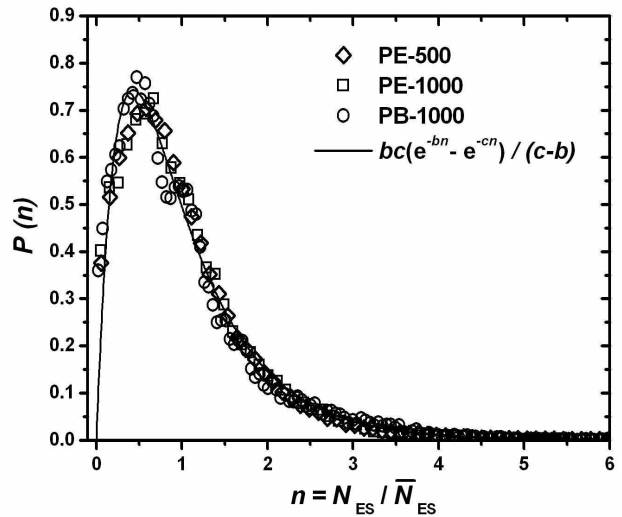}\end{center}

\caption{\label{cap:Ne}Normalized distribution of the reduced monomer distance
between entanglements for the polymers of Table \ref{table}.}
\end{figure}

In Fig.~\ref{cap:Ne} we present the \emph{normalized} distribution
$P(n)$ of the reduced number of monomers $N_{{\rm ES}}$ in an ES,
$n=N{\rm _{{\rm ES}}}/\overline{N}_{{\rm ES}}$. The data from both
materials superimpose on each other, falling on a master curve of
the form $P(n)=\frac{bc}{c-b}(e^{-bn}-e^{-cn})$, with fitted values
$b=1.30$, $c=3.78$. In view of the large difference in $\overline{N}_{{\rm ES}},\: p$
between PE and PB, the collapse of data suggests a universal character
of $P(n)$ for linear polymers. The distribution
is very broad with an exponential tail and in contrast with the tube
model implies a non-uniform network mesh and large fluctuations in
the number of monomers of an ES. The origin of such fluctuations \cite{Rubi}
and their effect on rheological properties have recently been examined
\cite{Greco,Schieb}. Schieber \cite{Schieb} considered a Gaussian chain
with a constant number of Kuhn segments in contact with a `bath' of
entanglements. By fixing the corresponding chemical potential, he
came up with an exponential distribution for the number of Kuhn segments
in an ES. Stochastically, this distribution can be generated \cite{Schieb} by marching
from one chain end to the other and placing entanglements on Kuhn
segments according to a Poisson process \cite{Cox} of `entanglement
events'. In our case the exponential tail of $P(n)$ is the result
of a similar stochastic process evolving on the \emph{monomer sequence
space} of a chain (see below). However, in a Poisson process successive
events can come arbitrarily close, while it is meaningful to assume
that an entanglement requires a certain number of successive monomers
to develop. This creates an \textit{effective repulsion} between successive
entanglements in the \textit{\emph{monomer sequence space}} of a chain,
which will become prominent close to and below the number of monomers
of the Kuhn segment of the atomistic chain. This is evidenced in Fig.
\ref{cap:Ne} by the downturn and vanishing trend of $P(n)$ data
as $n\rightarrow0$.

A \emph{stochastic interpretation} of our results can be given in
terms of a \emph{renewal process} \cite{Cox}, which is the generalization
of a Poisson process. In such a process, successive events are still
identically, independently distributed, though with a distribution
that is not exponential, cf. the normalized $P(n)$ presented here.
Moreover, $P(n)$ is the convolution of two exponential distributions,
namely $be^{-bn}$ and $ce^{-cn}$, and can be interpreted \cite{Cox}
as the result of two (uncorrelated) alternating Poisson processes
with rates $b,\, c,$ evolving on \emph{the monomer sequence
space} of a chain. The first process, with rate $c$, does not result
in observable events \cite{Cox}. What it does is to stochastically
create an unentangled monomer sequence in front of an entanglement,
corresponding to the mentioned repulsion. The second process, with
rate $b$, takes over when an unentangled sequence has been created
and places an entanglement in one of the following monomers.

\begin{figure}[t]
\begin{center}\includegraphics[%
  clip,
  scale=0.42]{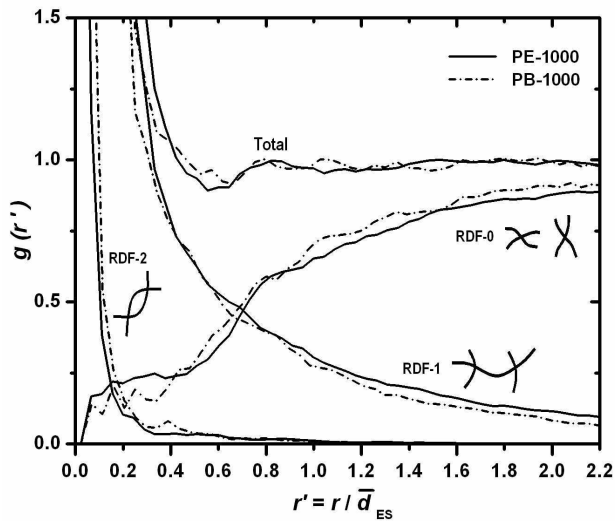}\end{center}

\caption{\label{cap:Rdf}Total and partial RDFs of PE-1000 (solid) and PB-1000
(dashed). Partial RDFs depict correlations between different pair
types of binary entanglements. RDF-0 curves correspond to entanglements
that are \emph{`spatial neighbors'}, i.e., all four chains passing
through them are different. RDF-1, RDF-2 curves correspond to entanglements
\emph{at any distance} that are \emph{`topological neighbors'}, i.e.,
connected by one or two common chains, respectively, passing through
them. Each case is sketched in the figure.}
\end{figure}

Finally, in Fig. \ref{cap:Rdf} we present the total and partial radial
distribution functions (RDFs) of entanglements versus $r'=r/\overline{d}_{{\rm ES}}$,
the spatial distance $r$ reduced by the average network mesh length.
Apart from some minor differences, the curves for PE, PB superimpose
on each other. Therefore, the spatial network nodal correlations stemming
from the underlying topologies of the PE, PB melts are similar when
scaled accordingly. The presence of very small network strands, which
connect first topological neighbors (see Fig. \ref{cap:Rdf}), is
responsible for the effective \emph{spatial attraction} evidenced
in the total RDF for $r'\lesssim0.4$, and separately in RDF-1,2.
The fact that RDF-0 curves are below RDF-1 curves for $r'\lesssim0.7$
indicates that in the region around and near an entanglement topological
neighbors persist and spatial neighbors cannot easily penetrate. The
latter populate primarily the region beyond $r'\sim0.7$ around a
network node, where RDF-0 curves overpass RDF-1 curves. RDF-2 curves
fall rapidly with $r'$ and correspond to pairs of chains intertwined
through two entanglements. Intertwining results in closely spaced
entanglements and is highly improbable at large $r'$. The total RDF
implies that in the range $r'\gtrsim0.8$ the network structure can
be described as an \emph{ideal dilute gas} of entanglements
\cite{Kramer}, with \emph{short range correlations} occurring at
smaller $r'$.

In summary, common organization properties of topological
constraints in polymer melts have been revealed through a 
statistical analysis of primitive paths. We are grateful to 
N. Karayannis, V. Mavrantzas and P. Gestoso for
providing the atomistic polymer samples. This work is funded by the
Dutch Polymer Institute.


\begin{thebibliography}{10}
\bibitem{foot1}We note that categorizing the chains into subsets and reducing
the subsets sequentially in arbitrary order leads practically to the
same network structure.
\end{thebibliography}
\end{document}